\documentclass[runningheads]{llncs}
\usepackage{graphicx}
\usepackage{array,multirow}

\begin{document}
\title{A Compact Representation of Histopathology Images using Digital Stain Separation \& Frequency-Based Encoded Local Projections\thanks{This research has been supported in part by a Natural Sciences and Engineering Research Council of Canada (NSERC) Doctoral Scholarship (AKC).}}
\titlerunning{Representation of Histopathology Images by Stain Separation \& F-ELP}
%
\author{Alison K. Cheeseman\inst{1} \and
Hamid Tizhoosh\inst{2} 
\and
Edward R. Vrscay\inst{1}}
\authorrunning{A.K. Cheeseman et al.}
%
\institute{
Department of Applied Mathematics, Faculty of Mathematics,
University of Waterloo, Waterloo, Ontario, Canada N2L 3G1\\
\email{alison.cheeseman@uwaterloo.ca, ervrscay@uwaterloo.ca}
\and
Kimia Lab, University of Waterloo, Waterloo, Ontario, Canada N2L 3G1\\
\email{hamid.tizhoosh@uwaterloo.ca}
}
\maketitle              
\begin{abstract}
In recent years, histopathology images have been increasingly used as a diagnostic tool in the medical field. The process of accurately diagnosing a biopsy sample requires significant expertise in the field, and as such can be time-consuming and is prone to uncertainty and error. With the advent of digital pathology, using image recognition systems to highlight problem areas or locate similar images can aid pathologists in making quick and accurate diagnoses. In this paper, we specifically consider the encoded local projections (ELP) algorithm, which has previously shown some success as a tool for classification and recognition of histopathology images. We build on the success of the ELP algorithm as a means for image classification and recognition by proposing a modified algorithm which captures the local frequency information of the image. The proposed algorithm estimates local frequencies by quantifying the changes in multiple projections in local windows of greyscale images. By doing so we remove the need to store the full projections, thus significantly reducing the histogram size, and decreasing computation time for image retrieval and classification tasks. Furthermore, we investigate the effectiveness of applying our method to histopathology images which have been digitally separated into their hematoxylin and eosin stain components. The proposed algorithm is tested on the publicly available invasive ductal carcinoma (IDC) data set. The histograms are used to train an SVM to classify the data. The experiments showed that the proposed method outperforms the original ELP algorithm in image retrieval tasks. On classification tasks, the results are found to be comparable to state-of-the-art deep learning methods and better than many handcrafted features from the literature.
\end{abstract}

\section{Introduction}

Histopathology, the examination of tissue under a microscope to study biological structures as they relate to disease manifestation, has recently attracted a lot of interest from the medical imaging research community. With the introduction of whole slide digital scanners, histopathology slides can now be digitized and stored in a digital form. As a result, it is now possible to apply computer-aided diagnosis and image analysis algorithms to the emerging field of digital histopathology~\cite{ref_gurcan09}. Content-based image retrieval (CBIR) and image classification are two important components of computer-aided image analysis which we consider in this paper. In a classification approach, the objective is to classify each image as belonging to a disease category. Image retrieval involves finding images which share the same visual characteristics as the query image. The identification and analysis of similar images can assist pathologists in quickly and accurately obtaining a diagnosis by providing a baseline for comparison.

As a result of the extremely large size of digital histopathology images, it is desirable to generate compact image descriptors for both retrieval and classification tasks. A large number of well-known image descriptors already exist, however the different requirements of a new application make constant innovation necessary. This becomes particularly important in a field where trained feature extraction algorithms, such as deep networks, may not always be feasible. Deep networks require massive volumes of labelled data for optimal training, yet large (and balanced) data sets are not always available in the medical field. This is especially true when we consider digital histopathology, as obtaining ground truth annotations is time-consuming and requires expert knowledge. Handcrafted image descriptors, such as the well-known local binary patterns (LBP)~\cite{ref_lowe04}, scale-invariant feature transform (SIFT)~\cite{ref_ahonen06}, and histogram of oriented gradients (HOG)~\cite{ref_dalal05} get around this issue by incorporating expert knowledge directly into their design without requiring any training data. Such descriptors and their successors have been quite successful in a range of diverse imaging applications~\cite{ref_tizhoosh18}. More recently, a projection-based histogram descriptor was proposed in~\cite{ref_tizhoosh18} specifically for the application of CBIR and classification of medical images. The ELP image descriptor has been very successful thus far on histopathology images, outperforming many well-known handcrafted features, and even outperforming some deep features generated using a convolutional neural network (CNN) when applied to medical imaging applications~\cite{ref_tizhoosh18}.

In this paper we build upon the ELP descriptor, a dense-sampling method introduced in~\cite{ref_tizhoosh18} and propose a frequency-based ELP (F-ELP) descriptor which  captures the local frequency information of the image. Instead of storing entire projections, as in the ELP method, our proposed method quantifies the number of changes in each projection and uses this as an estimate of local frequency. While the original ELP method results in large histograms, the size of our F-ELP histograms has linear dependence on the local window size. The compact nature of our descriptor is desirable from the perspective of both memory usage for storage of descriptors and computation requirements when applied to image retrieval and classification type tasks. In addition to the introduction of our novel histogram representation, we also discuss the use of digital stain separation to improve the performance of our descriptor.

 We test the performance of the proposed F-ELP descriptor on both image retrieval and image classification tasks. The publicly available invasive ductal carcinoma (IDC) dataset is used to evaluate the performance of our method and for comparison to state-of-the-art results from the literature.
 
 \section{The Proposed Method}
 
 Our proposed method involves two main innovations, which are described in more detail here. First, we introduce our proposed image descriptor and discuss how it differs from the ELP descriptor in~\cite{ref_tizhoosh18}. We then introduce the idea of separating the histopathology images into their histochemical stain components to generate more meaningful image descriptors. 

\subsection{Frequency-based ELP (F-ELP)}

The ELP image descriptor is a dense-sampling method which encodes the gradient changes of multiple Radon projections in small local windows of the image. To maintain some level of robustness to rotation, a dominant angle is determined for each local neighbourhood using the image gradient. The dominant, or anchor, angle is then used to anchor the projections in that window so that the end result does not depend on image orientation. Projection gradients are encoded using the \textit{MinMax}~\cite{ref_tizhoosh18} method to generate a binary number which is then used to build a histogram. It should be noted here that in~\cite{ref_tizhoosh18} an alternate method of computing the anchor angle based on the overall Radon sinogram is initially proposed. However, when applying their method to a pathology data set, the authors choose to save time by approximating the anchor angle computation by the median of the image gradient directions. In order to have a fair comparison, we consider this particular implementation of the ELP descriptor in this work.

The computation of our proposed F-ELP descriptor follows the same overall steps as the ELP descriptor, with some modifications along the way to improve rotational invariance, reduce sensitivity to shifts in the image and reduce redundancy by encoding only the frequency information from each projection. We describe each step in detail as follows, highlighting where our method differs from the ELP method.

\subsubsection{Identify local windows:}

The first step is to identify a set of small local windows for processing. Here, our method does not differ at all from the original ELP method. Since we are interested in finding projections which uniquely describe the patterns/textures in local neighbourhoods, we only consider regions which are sufficiently non-homogeneous so as to ensure projections contain something of interest. We let $\mathbf{W}$ denote a local window of size $n \times n$ and calculate the homogeneity, $H$ of each window according to
\begin{equation}
    H = 1 - \frac{1}{2^{n_{bits}}} \sqrt{\sum_i\sum_j \left(\mathbf{W}_{ij} - m \right) ^2},
\end{equation}
where $m$ denotes the median pixel value of $\mathbf{W}$ and $n_{bits}$ is the number of bits used to encode the image. A threshold is used to eliminate any windows with high homogeneity.

\subsubsection{Determine the anchor angle, $\theta^*$:}

In order for our descriptor to be rotationally invariant, we seek a unique angle in each window by which to ``anchor" our projections. We do so by computing the image gradient, binning the gradient directions into one degree intervals and selecting $\theta^*$ to be the mode (most frequently occurring) of the gradient directions. Our approach differs just slightly from the original method, in that we choose to use the mode instead of the median to find the average angle. We do so as the median is not invariant under circular shifts (i.e. angular rotations), whereas the mode is, so long as there is one unique angle which occurs at the highest frequency (i.e. a clearly dominant direction in the window). 

\subsubsection{Compute the projection along $\theta^*$:} As in the ELP method  we compute projections using the Radon transform, which is given by
\begin{equation}
    R(\rho,\theta) = \int_{-\infty}^{\infty}\int_{-\infty}^{\infty}f(x,y)\delta(\rho-x\cos\theta-y\sin\theta)dxdy,
\end{equation}
where $\delta(\cdot)$ is the Dirac delta function. We extract the projection $\mathbf{p}_{\theta^*}$ by taking the Radon transform along parallel lines $\rho$ for the fixed anchor angle $\theta^*$. 

\subsubsection{Encode projections and create histogram:}

It is in the encoding of the projections where our algorithm differs the most notably from the ELP method. Instead of encoding the entire gradient of each projection, we quantify the gradient changes in the projection vector and use this to build our histogram. The benefits of this modification are two-fold. Primarily, we remove the storage overhead of encoding entire projections, and instead just capture the general trend (low or high frequency) of the projections along each direction, resulting in much smaller histograms which still perform very well. Our proposed method also avoids the use of a binary encoding to capture the projections. This is beneficial as the binary encoding used by the ELP is very sensitive to small shifts in the projection, i.e. a change in one binary bit can lead to a very large difference in the resulting histogram. On the other hand, when the local projection frequency changes, the resulting change in the F-ELP histogram reflects the size of the frequency change.

Given a projection vector $\mathbf{p}$ of length $n$ and its derivative $\mathbf{p}'$, we compute the following quantized encoding of the derivative,
\begin{equation}
\label{eq_ternary}
    \mathbf{q}(i) = \left\{
\begin{array}{ll}
      0 & \mbox{\ if\ } \mathbf{p}'(i) \leq -T \\
      1 & \mbox{\ if\ } |\mathbf{p}'(i)| < T \\
      2 & \mbox{\ if\ } \mathbf{p}'(i) \geq T. \\
\end{array} 
\right.
\end{equation}
The three levels given in (\ref{eq_ternary}) indicate regions where the projection, $\mathbf{p}$, is decreasing, nearly constant (we use a small threshold, $T$, here to ignore small fluctuations), and increasing, respectively. Next, we count the number of transitions in $\mathbf{q}$ to get our estimate of local frequency which will be an integer value, $d$ which satisfies $0 \leq d \leq n-2$. Once we have $d$ we can increment the histogram $\mathbf{h}(d)$.

\begin{figure}
\centering
\includegraphics[width=\textwidth]{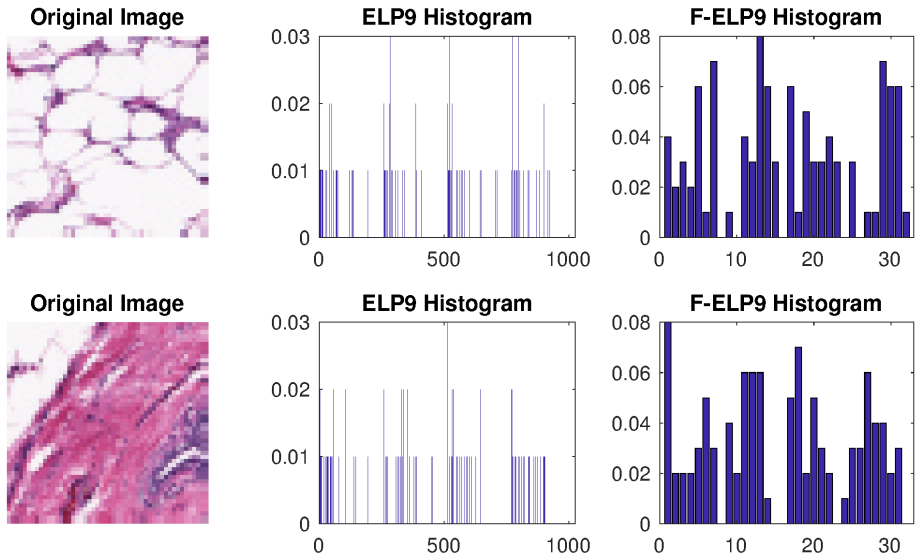}
\caption{Sample histograms generated using the ELP and F-ELP methods with a window size of $n=9$.} \label{fig_histplot}
\end{figure}

Similar to the original ELP descriptor, we obtain more information by computing three additional projections relative to our anchor angle $\theta^*$. These are equidistant projections, given by $\Theta = \left\{\theta^*,\theta^*+\pi/4,\theta^*+\pi/2,\theta^*+3\pi/4 \right\}$. For each additional angle, the projections are computed and encoded in the same manner.  The final histogram is generated by concatenating all four histograms into one longer histogram.

Figure~\ref{fig_histplot} shows an example of both the ELP and F-ELP descriptors for two sample images from the IDC dataset which contain somewhat different textures. In both cases, the histograms have been normalized according to the $L1$-norm. We see that the F-ELP descriptor, although it has less bins, appears to show a more varied distribution. When looking at the ELP histograms, we observe that the distribution is similar for both images, with many bins empty. This indicates that there is some redundancy in this image representation which we try to remove using the F-ELP method.

\subsection{Digital Stain Separation}

Prior to imaging, histology slides are stained to enhance the detail in tissues and cells. The most common stain protocol used in practice is hematoxylin and eosin (H\&E), where hematoxylin components stain cell nuclei blue, and eosin stains other structures varying shades of red and pink~\cite{ref_ruifrok01}. The colours which appear in a slide and the size, shape and frequency at which they appear are all relevant factors a pathologist might assess when making a diagnosis. For this reason, we consider separating the input images into two components which give the amount of each stain at each pixel and computing a descriptor for each component.
\begin{figure}
\centering
\includegraphics[width=\textwidth]{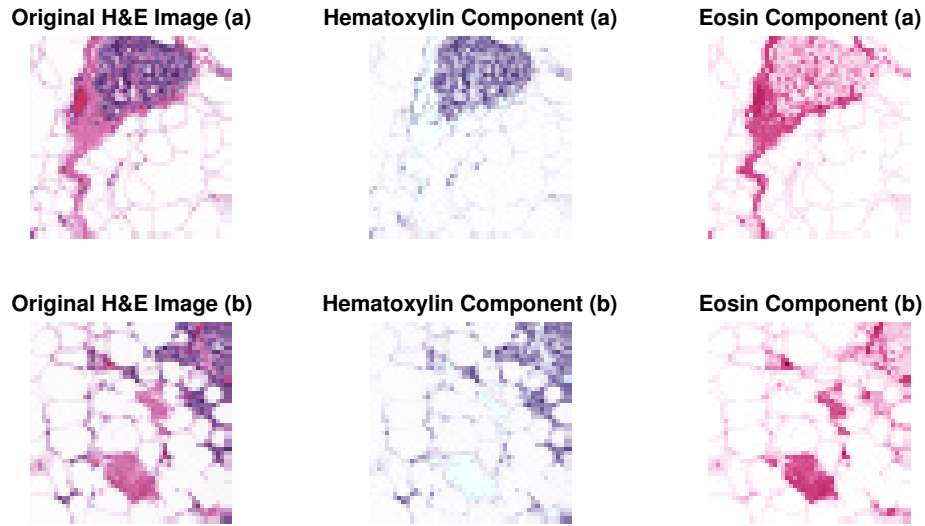}
\caption{Sample images from the IDC dataset showing the hematoxylin and eosin stane components after applying the stain separation algorithm from~\cite{ref_mccann14}.} \label{fig_stainsep}
\end{figure}

A number of methods already exist in the literature for digital stain separation of H\&E slides that perform quite well. Although their intended usage is for stain normalization to control for variation in stain intensities and colours, these same methods are suitable for our purposes. In this paper we adopt the method proposed in~\cite{ref_mccann14}, an extension of the wedge finding method from ~\cite{ref_macenko09}. Unlike some previous methods for stain separation~\cite{ref_ruifrok01}, this method does not require any calibration or knowledge of the exact stain colours, instead it works by using the available image data to estimate an H\&E basis. Given that our image descriptor should ultimately be applied to data from multiple sources, this is an important feature of the stain separation algorithm. Figure~\ref{fig_stainsep} shows two examples of image patches before and after stain separation is applied. We can see that the algorithm does a good job of isolating the hematoxylin (blue/purple) component from the eosin (pink/red) component, even in regions where both stains contribute to the overall pixel colour.

Given the resulting stain separated image components, we proceed as described above to compute our F-ELP image descriptor, simply computing the F-ELP histogram for each component of the image separately. This results in two histograms, $\mathbf{h}_H$ and $\mathbf{h}_E$ which are concatenated to form the final longer histogram $\mathbf{h} = [\mathbf{h}_H\ \mathbf{h}_E]$.

\section{Dataset \& Image Preprocessing}

We have used the publicly available IDC data set to test our method in both image retrieval and image classification. The dataset consists of digitized breast cancer slides from 162 patients diagnosed with IDC at the University of Pennsylvania Hospital and the Cancer Institute of New Jersey~\cite{ref_cruz-roa14}. Each slide was digitized at 40x magnification and downsampled to a resolution of 4 $\mu$m/pixel. The supplied data was randomly split into three different subsets of 84 patients for training, 29 for validation and 49 test cases for final evaluation. The dataset provides each WSI split into image patches which are 50px x 50px in size. Ground truth annotation regarding the presence of IDC in each patch was obtained by manual delineation of cancer regions performed by expert pathologists. 

For each image patch, we computed the F-ELP descriptor using the method described in the previous section. For these experiments we implement the algorithm with a threshold $T=0.08$, determined by observation of the data, and a homogeneity threshold of $1$, rejecting only completely homogeneous windows within each patch. In a future work, it would be beneficial to optimize these parameters and carry out some form of sensitivity analysis.

Since the image patches are quite small, each patch may not contain the presence of both hematoxylin and eosin. For better results, we use the entire whole slide image (WSI) to perform the stain separation and then split the image back into 50 x 50 patches to compute individual histograms. The stain separation algorithm we used assumes two stain components (H\&E in our case) exist in the image, however some images are observed to have significant discoloration, such as large dark blue patches, and the introduction of other colours not caused by H\&E staining. The prevalence of such artefacts negatively impacts the ability of the stain separation algorithm to provide good results for some patients, so we remove them by searching for images which have minimal variation in the RGB channels across the entire image. A total of 686 images were flagged and removed from the total data set, all of which contain significant artefacts/discoloration. 

To evaluate the performance of our method in both image retrieval and classification tasks we have used the provided test data set consisting of 49 patients as our input data. For consistency with previous works we use both the balanced accuracy (BAC) and F-measure (F1) as performance metrics, which are defined as follows~\cite{ref_cruz-roa14}:
\begin{equation}
    \mbox{BAC} = \frac{\mbox{Sen}+\mbox{Spc}}{2}
\end{equation}
\begin{equation}
    F1 = \frac{2 \cdot \mbox{Pr} \cdot \mbox{Rc}}{\mbox{Pr}+\mbox{Rc}},
\end{equation}
where Sen is sensitivity, Spc is specificity, Pr is precision and Rc, recall.

\section{Experimental Results}

In this section we present the results of using our proposed F-ELP descriptor for both image retrieval and classification of the IDC dataset. For each task we compare our method to relevant methods from the literature.

\subsection{Image Retrieval}

In order to evaluate the image retrieval performance of our descriptor we implement the k-Nearest Neighbours (kNN) algorithm in MATLAB with the F-ELP histograms as inputs. The kNN algorithm searches through the training data partition and classifies each image based on the class of its $k$ nearest neighbours. Since there is no exact metric to quantitatively test image retrieval performance, we evaluate the expressiveness of the ELP-based descriptors for image retrieval tasks based on the accuracy of classification using kNN. In this work, we test the kNN algorithm using three different values for $k$ ($k=1,3$ and $5$). Four different distance metrics  were used to determine the nearest neighbours, including the commonly used $L_1$, $L_2$ and cosine distances. We also used the Hutchinson (or Monge-Kantorovich) distance~\cite{ref_mendivil17}, a metric between two probability measures, as it is considered to be a good measure of distance between histograms. In the finite one-dimensional case, the Hutchinson distance can be computed in linear-time using the method from~\cite{ref_molter91}.
\begin{table}
\centering
\caption{F1 \& BAC results for image retrieval ($k=1$) on the IDC dataset}\label{tab_search1}
\begin{tabular}{|c|c|c|c|c|c|c|c|c|}
\hline
\multirow{2}{*}{Method} & \multicolumn{2}{c|}{L1} & \multicolumn{2}{c|}{L2} & \multicolumn{2}{c|}{Cosine} & \multicolumn{2}{c|}{Hutchinson}\\\cline{2-9}
& F1 & BAC & F1 & BAC & F1 & BAC & F1 & BAC\\\hline
ELP9 & 0.3072 & 0.5616 & 0.3728 & 0.5842 & 0.2965 & 0.5594 & \textbf{0.3985} & \textbf{0.5904}\\
ELP9 + SS & 0.3004 & 0.5654 & 0.4339 & 0.6167 & 0.2688 & 0.5574 & \textbf{0.4527} & \textbf{0.6236}\\
F-ELP9 & 0.4177 & 0.6008 & 0.4179 & 0.6015 & \textbf{0.4200} & \textbf{0.6022} & 0.4189 & 0.6025\\
F-ELP9 + SS & 0.5489 & 0.6881 & 0.5486 & 0.6879 & \textbf{0.5504} & \textbf{0.6891} & 0.5472 & 0.6869\\
F-ELP11 & \textbf{0.3969} & \textbf{0.5865} & 0.3947 & 0.5855 & 0.3954 & 0.5863 & 0.3792 & 0.5774\\
F-ELP11 + SS & 0.5237 & 0.6707 & 0.5173 & 0.6666 & 0.5185 & 0.6673 & \textbf{0.5347} & \textbf{0.6784}\\
\hline
\end{tabular}
\end{table}

\begin{table}
\centering
\caption{F1 \& BAC results for image retrieval ($k=3$) on the IDC dataset}\label{tab_search3}
\begin{tabular}{|c|c|c|c|c|c|c|c|c|}
\hline
\multirow{2}{*}{Method} & \multicolumn{2}{c|}{L1} & \multicolumn{2}{c|}{L2} & \multicolumn{2}{c|}{Cosine} & \multicolumn{2}{c|}{Hutchinson}\\\cline{2-9}
& F1 & BAC & F1 & BAC & F1 & BAC & F1 & BAC\\\hline
ELP9 & 0.4009 & 0.6092 & 0.4837 & 0.6432 & 0.3807 & 0.6015 & \textbf{0.5117} & \textbf{0.6523}\\
ELP9 + SS & 0.3662 & 0.5976 & 0.5372 & 0.6743 & 0.3240 & 0.5830 & \textbf{0.5609} & \textbf{0.6833}\\
F-ELP9 & 0.5106 & 0.6514 & 0.5100 & 0.6515 & \textbf{0.5122} & \textbf{0.6526} & 0.5056 & 0.6490\\
F-ELP9 +SS & 0.6267 & 0.7350 & 0.6251 & 0.7339 & 0.6257 & 0.7345 & \textbf{0.6309} & \textbf{0.7375}\\
F-ELP11 & 0.5010 & 0.6420 & 0.5034 & 0.6443 & \textbf{0.5043} & \textbf{0.6450} & 0.4879 & 0.6355\\
F-ELP11 + SS & 0.6117 & 0.7223 & 0.6003 & 0.7149 & 0.6053 & 0.7184 & \textbf{0.6190} & \textbf{0.7279}\\
\hline
\end{tabular}
\end{table}

\begin{table}
\centering
\caption{F1 \& BAC results for image retrieval ($k=5$) on the IDC dataset}\label{tab_search5}
\begin{tabular}{|c|c|c|c|c|c|c|c|c|}
\hline
\multirow{2}{*}{Method} & \multicolumn{2}{c|}{L1} & \multicolumn{2}{c|}{L2} & \multicolumn{2}{c|}{Cosine} & \multicolumn{2}{c|}{Hutchinson}\\\cline{2-9}
& F1 & BAC & F1 & BAC & F1 & BAC & F1 & BAC\\\hline
ELP9 & 0.4138 & 0.6159 & 0.5057 & 0.6559 & 0.3904 & 0.6064 & \textbf{0.5405} & \textbf{0.6699}\\
ELP9 + SS & 0.3599 & 0.5948 & 0.5563 & 0.6861 & 0.3142 & 0.5786 & \textbf{0.5897} & \textbf{0.7016}\\
F-ELP9 & 0.5345 & 0.6659 & 0.5314 & 0.6645 & \textbf{0.5364} & \textbf{0.6675} & 0.5284 & 0.6629\\
F-ELP9 +SS & 0.6492 & 0.7505 & 0.6485 & 0.7498 & 0.6474 & 0.7494 & \textbf{0.6521} & \textbf{0.7519}\\
F-ELP11 & 0.5294 & 0.6591 & \textbf{0.5303} & \textbf{0.6603} & 0.5282 & 0.6595 & 0.5155 & 0.6519\\
F-ELP11 + SS & 0.6381 & 0.7398 & 0.6306 & 0.7349 & 0.6309 & 0.7351 & \textbf{0.6427} & \textbf{0.7437}\\
\hline
\end{tabular}
\end{table}

For comparison purposes, we implement the image retrieval algorithm with both the original ELP descriptor and our F-ELP descriptor as inputs. Both descriptors are implemented with and without stain separation (SS) of the image data. The ELP descriptor was designed using a window size of $n=9$, which is what we implement here using code obtained from the authors of~\cite{ref_tizhoosh18}. This results in histograms of length $1024$ (without SS) and $2048$ (with SS). Since the F-ELP descriptor is much shorter in length, in addition to $n=9$, we also test a larger window size, $n=11$. Even with this larger window size, the maximum histogram length for the F-ELP is just $80$ bins. Tables \ref{tab_search1}, \ref{tab_search3} and \ref{tab_search5} summarize the results of our comparison for the three values of $k$ implemented. For each method, the best performance across all distance metrics is highlighted in bold. We observe that as $k$ is increased the F-measure and balanced accuracy improve for all methods. Based on these results we expect that implementations of the kNN algorithm with even larger values of $k$ may yield even further improvements in accuracy measures, however in this work we are primarily concerned with the comparison between descriptors.

From the above results we observe that for both descriptors, the use of stain separation to generate histograms improves performance. In particular, we see a significant improvement in accuracy scores when we apply stain separation to the F-ELP descriptor. With stain separation, our proposed method significantly outperforms the ELP descriptor, all with much shorter histograms. In general, we find that the choice of distance function used does not seem to have a significant effect on the performance of the image retrieval algorithm, with accuracy scores being fairly similar across all distance functions. 

\subsection{Image Classification}

To evaluate image classification performance we used our F-ELP descriptor as input to train a support vector machine (SVM), a popular classification algorithm. We used the provided training data partition and the {\tt fitcsvm} function in MATLAB to train an SVM to classify each image as containing IDC cells or not. The SVM hyperparameters were optimized using a Bayesian optimization routine in MATLAB and two kernel functions, linear and Gaussian, were tested. As in the previous section, we implemented our method with two different window sizes and with/without stain separation. In Table~\ref{tab_class1} the classification results obtained with the optimal hyperparameters are recorded for each implementation of the F-ELP.

From the results in Table~\ref{tab_class1} we can see that using stain separation to generate our descriptor provides a significant improvement in classification performance for both window sizes that were tested.
\begin{table}
\centering
\caption{F1 \& BAC results for classification of the IDC dataset using F-ELP with and without stain separation}
\label{tab_class1}
\begin{tabular}{|c|c|c|c|c|c|c|c|c|}
\hline
Method &  F1 & BAC\\\hline
F-ELP9 & 0.4048 & 0.6174\\
\textbf{F-ELP9 + SS} & \textbf{0.7182} & \textbf{0.8076}\\
F-ELP11 & 0.3385 & 0.5911\\
F-ELP11 + SS & 0.6715 & 0.7665\\
\hline
\end{tabular}
\end{table}

From Table~\ref{tab_class1} we can also observe that classification performance is significantly better when we use the smaller window size. The optimal window size for a given task is likely determined by a number of factors, including image magnification and scale of the textures and patterns which are relevant to the particular classification task at hand. We did not test other window sizes at this time, however one could perform further testing to determine an optimal window size for the given application.

We now compare the performance of our best proposed method to the most recent results from the literature for classification of the IDC data, including handcrafted features and deep learning approaches. In Table~\ref{tab_class2} we see that our F-ELP descriptor outperforms many of the state-of-the-art handcrafted features, achieving accuracy levels that are more comparable to those of recent deep learning approaches, such as a CNN from~\cite{ref_cruz-roa14} and Alexnet, as implemented in~\cite{ref_janowczyk16}.
\begin{table}
\centering
\caption{F1 \& BAC results for classification of the IDC dataset}
\label{tab_class2}
\begin{tabular}{|c|c|c|c|c|c|c|c|c|}
\hline
Method &  F1 & BAC\\\hline
Alexnet,resize~\cite{ref_janowczyk16} & 0.7648 & 0.8468\\
CNN~\cite{ref_cruz-roa14} & 0.7180 & 0.8423\\
\textbf{F-ELP9 + SS} & \textbf{0.7182} & \textbf{0.8076}\\
Fuzzy Color Histogram~\cite{ref_cruz-roa14} & 0.6753 & 0.7874\\
RGB Histogram~\cite{ref_cruz-roa14} & 0.6664 & 0.7724\\
Gray Histogram~\cite{ref_cruz-roa14} & 0.6031 & 0.7337\\
JPEG Coefficient Histogram~\cite{ref_cruz-roa14} & 0.5758 & 0.7126\\
M7 Edge Histogram~\cite{ref_cruz-roa14} & 0.485 & 0.6979\\
Nuclear Textural~\cite{ref_cruz-roa14} & 0.3915 & 0.6199\\
LBP~\cite{ref_cruz-roa14} & 0.3518 & 0.6048\\
Nuclear Architectural~\cite{ref_cruz-roa14} & 0.3472 & 0.6009\\
HSV Color Histogram~\cite{ref_cruz-roa14} & 0.3446 & 0.6022\\
\hline
\end{tabular}
\end{table}

Although we do not achieve quite the same accuracy level of deep-learning approaches at this time, we find these results very encouraging. Using our F-ELP method with digital stain separation, we are able to achieve comparable accuracy with a much simpler approach. Our proposed method does not require any training data to generate the descriptors as they are handcrafted, meaning that a data set can be encoded using our method fairly quickly.

\section{Conclusion}

In this paper, we have introduced a new frequency-based descriptor for digital histopathology images. The proposed descriptor, F-ELP, is a histogram descriptor which estimates directional frequency information from local image patches. The F-ELP method outperforms its successor, the ELP descriptor, in image retrieval tasks, while requiring less storage overhead and shorter computation times due to its compact nature.

When compared to the state-of-the-art from the literature, our method outperforms a number of popular handcrafted features as an image classifier. We achieve classification results which are comparable to those of deep-learning methods. Our method achieves these results with a very compact representation that does not require large amounts of training data to generate. Our descriptor is also physically meaningful, being based on estimates of local frequency, and thus has the potential to be interpreted by medical experts.

%
%
%
\bibliographystyle{splncs04}
\bibliography{references}

\begin{thebibliography}{10}
\providecommand{\url}[1]{\texttt{#1}}
\providecommand{\urlprefix}{URL }
\providecommand{\doi}[1]{https://doi.org/#1}

\bibitem{ref_ahonen06}
Ahonen, T., Hadid, A., Pietikainen, M.: Face description with local binary
  patterns. IEEE Trans. Pattern Anal. Mach. Intell.  \textbf{28}(12),
  2037--2041 (2006)

\bibitem{ref_cruz-roa14}
Cruz-Roa, A., Basavanhally, A., Gonzalez, F., Gilmore, H., Feldman, M.,
  Ganesan, S., Shih, N., Tomaszewski, J., Madabhushi, A.: Automatic detection
  of invasive ductal carcinoma in whole slide images with convolutional neural
  networks. Progress in Biomedical Optics and Imaging - Proceedings of SPIE
  \textbf{9041} (2014)

\bibitem{ref_dalal05}
Dalal, N., Triggs, B.: Histograms of oriented gradients for human detection.
  In: 2005 IEEE Computer Society Conference on Computer Vision and Pattern
  Recognition. pp. 886--893. San Diego, CA, USA (2005)

\bibitem{ref_gurcan09}
Gurcan, M.N., Boucheron, L.E., Can, A., Madabhushi, A., Rajpoot, N.M., Yener,
  B.: Histopathological image analysis: A review. IEEE Rev. in Biomed. Eng.
  \textbf{2}(2),  147--171 (2009)

\bibitem{ref_janowczyk16}
Janowczyk, A., Madabhushi, A.: Deep learning for digital pathology image
  analysis: A comprehensive tutorial with selected use cases. J. Pathol.
  Inform.  \textbf{7}(29) (2016)

\bibitem{ref_lowe04}
Lowe, D.G.: Distinctive image features from scale-invariant keypoints.
  International Journal of Computer Vision  \textbf{60}(2),  91--110 (2004)

\bibitem{ref_macenko09}
Macenko, M., Niethammer, M., Marron, J.S., Borland, D., Woosley, J.T., Guan,
  X., Schmitt, C., Thomas, N.E.: A method for normalizing histology slides for
  quantitative analysis. In: Proceedings of IEEE Int. Symp. Biomed. Imag. pp.
  1107--1110. Chicago,IL (2009)

\bibitem{ref_mccann14}
McCann, M.T., Majumdar, J., Peng, C., Castro, C.A., Kovacevic, J.: Algorithm
  and benchmark dataset for stain separation in histology images. In:
  Proceedings of 2014 IEEE International Conference on Image Processing (ICIP).
  pp. 3953--3957. Paris, France (2014)

\bibitem{ref_mendivil17}
Mendivil, F.: Computing the monge-kantorovich distance. Computational and
  Applied Mathematics  \textbf{36}(3),  1389--1402 (2017)

\bibitem{ref_molter91}
Molter, U., Brandt, J., Cabrelli, C.: An algorithm for the computation of the
  hutchinson distance. Information Processing Letters  \textbf{40}(2),
  113--117 (1991)

\bibitem{ref_ruifrok01}
Ruifrok, A.C., Johnston, D.A.: Quantification of histochemical staining by
  color deconvolution. Analytical and quantitative cytology and histology/T
  \textbf{23}(4),  291--299 (2001)

\bibitem{ref_tizhoosh18}
Tizhoosh, H.R., Babaie, M.: Representing medical images with encoded local
  projections. IEEE Trans. Biomed. Eng.  \textbf{65}(10),  2267--2277 (2018)

\end{thebibliography}
\end{document}